# Correlation of morphology and charge transport in poly(3,4-ethylenedioxythiophene)–polystyrenesulfonic acid (PEDOT–PSS) films

C.S. Suchand Sangeeth, Manu Jaiswal, Reghu Menon

*Department of Physics, Indian Institute of Science, Bangalore 560012, India*

Email: manujaiswal81@gmail.com, manu@physics.iisc.ernet.in

Abstract

A wide variation in the charge transport properties of poly(3,4-ethylenedioxythiophene)-poly(styrenesulfonate) (PEDOT-PSS) films is attributed to the degree of phase-segregation of the excess insulating polyanion. The results indicate that the charge transport in PEDOT-PSS can vary from hopping to critical regime of the metal-insulator transition, depending on the subtle details of morphology. The extent of electrical-connectivity in films, directly obtained from a temperature-dependent high-frequency transport study, indicates various limiting factors to the transport, which are correlated with the phase separation process. The low temperature magnetotransport further supports this morphology-dependent transport scenario.

PACS: 73.61.Ph, 72.80.Le, 73.43.Qt

## 1. Introduction

The current progress in organic/polymer optoelectronics and device technology relies significantly on the ability to tune the various device transport parameters by means of morphology, interface modification, controlling the extent of disorder, varying the charge carrier density and mobility. The performance of these devices is strongly interlinked with the morphology and structure of the constituting organic material [1]. Significant enhancement in performance has been achieved by tuning the morphology, notably in the case of polythiophene transistors [2] or polymer/methanofullerene blended organic solar cell [3]. Among the important conducting polymers for device applications is poly(3,4-ethylenedioxythiophene):poly(styrenesulfonate) (PEDOT-PSS), consisting of a conducting polythiophene derivative that is electrostatically bound to a PSS polyanion [4, 5]. Since this polymeric system simultaneously has high electrical conductivity, processibility, high stability, flexibility and good transparency, it has found large variety of applications including polymeric anodes for organic photovoltaics, [6, 7] light-emitting diodes, [8-11] flexible electrodes, [12] supercapacitors, [13] electrochromic devices, [14] field-effect transistors [4] and antistatic-coatings [15]. Usually, the appropriate combination of surface morphology and conductivity is required for various applications of PEDOT-PSS. The processing of PEDOT-PSS depends on a variety of factors including the fraction of PSS, fraction of solid content, particle size and the viscosity – which can lead to a wide range of morphological and electronic characteristics in the dried films. Antistatic coating applications require high conductivity for the PEDOT-PSS layer. In organic light-emitting diode (OLED) applications, the PEDOT-PSS layer serves as an efficient hole-injecting layer into the active semiconductor. This requires lower conduction for PEDOT-PSS but the interface between the active layer and polymer must be smooth in order to prevent micro-shorts leading to dark spots [4]. Furthermore, the surface roughness of highly conducting PEDOT-PSS (PH500 from H.C. Stark) was recently shown to be responsible for the reduced performance in polymer solar cells [16]. More recently, PEDOT-PSS (PH510 from H.C. Stark) has been utilized as highly conducting polymer anode in a high-efficiency white OLED [17]. In this context, a study of the correlation between morphology and conduction for PEDOT-PSS can address some of the issues involved in device applications.

Earlier it has been shown that the charge transport in an individual PEDOT-PSS nanowire is quasi one-dimensional in nature [18]. The processing of PEDOT-PSS film requires formation of complex with PSS which serves as charge balancing counter ion as well as causes the dispersion of PEDOT segments in aqueous solution. When processed with organic solvents such as glycerol and dimethyl sulfoxide (DMSO), the conductivity of the films is enhanced by two orders of magnitude [19]. This phenomenon was initially attributed to the screening effects of polar solvents on the Coulomb interaction between positively charged PEDOT grain and negatively charged PSS dopant [20] or to solvent-induced a conformational change in the PEDOT chain, which increases the inter-chain interaction [21]. Recent investigations indicate that this conductivity enhancement is due to morphological changes in the film structure and a phase-segregation process. PEDOT-PSS is known to form a phase-segregated material comprising highly-conducting PEDOT grains that are surrounded by excess weakly ionic-conducting PSS [22-28]. The anisotropic conductivity of PEDOT-PSS was recently correlated with the nanoscopic morphology in the phase-segregation scenario [29]. The polar solvent is known to enhance the extent of phase separation process [8]. Since the excess PSS is partly phase-separated, the effective insulation of the conducting PEDOT-PSS particles by the excess PSS can be considerably reduced, leading to better pathways for conduction and an overall increase in the connectivity of PEDOT-PSS network. Atomic force microscopy characterization has suggested the presence of PEDOT-rich and PSS-rich regions on the surface but a detailed correlation with the transport is still lacking [30].

While recent investigations have supported the picture of phase-segregated model for PEDOT-PSS, the extent to which the insulating PSS lamellae determine the transport needs a detailed investigation. In the present work, we present transport measurements at low temperatures, high magnetic fields and also high frequencies to investigate the network connectivity in the PEDOT-PSS system, and correlate with the film morphology. In particular, we observe that the limitation on conduction in the large-grain PEDOT-PSS arises not from the insulating PSS barriers but rather from the intrinsic disorder present within the PEDOT-rich regions itself.

2. Experimental details

The PEDOT- PSS films were prepared from two aqueous dispersions with trade names Baytron PH510 and P, obtained from H.C. Stark. The weight ratio of PEDOT to PSS was 1:2.5. The solution was filtered with 0.2 m membrane to remove traces of macroscopic aggregates and mixed with dimethyl sulfoxide (DMSO). The volume fraction of DMSO was 5 % for PH510 solution and 25 % for P solution. The glass substrates used were ~ 2 cm square and these were pre-cleaned by ultrasonication in acetone, triple distilled water and isopropyl alcohol for 30 min each. After stirring, the films were solution cast on glass substrate held at ~ 70 $^oC$ and annealed for 24 hours, and then peeled off after drying to obtain free standing films. Tapping-mode Atomic Force Microscopy (TM-AFM) measurements were performed on freshly prepared films, in Veeco Instruments/CPR2.

The transport measurements are performed by standard four-probe dc technique, and conductive carbon paint was used to make contacts (~ 1 mm apart) on the sample. For the conductance measurement, the current is sourced through Keithley 220 current source and the voltage measured through Keithley 2000 multimeter. The current values were typically in the range of 0.2 to 1 μA, chosen to avoid any sample heating at low temperatures. The sample is immersed in Helium vapor/liquid in a Janis variable-temperature cryogenic system equipped with a superconducting magnet. The measurements are performed in magnetic fields up to 11 Tesla and temperatures down to 1.3 K. The ac impedance is measured in the standard four-terminal configuration from 100 kHz to 10 MHz using Agilent-4285A LCR meter in a Janis continuous-flow cryogenic system down to 4.2 K. These measurements were further extended to lower frequency (40 Hz) using a lock-in amplifier (Stanford Research 830) relying on resistor-divider setup to obtain the sample conductance at low frequencies.

3. Results and discussion

The Tapping-mode Atomic Force Microscopy (TM-AFM) images of the films are shown in Figure 1 in the topography mode. The morphology of the PEDOT-PSS films reveals granular structures which can be assigned to the polymer nanoparticles of the dispersion solution, consistent with previous observations [7, 30]. The extent of aggregation of the PEDOT-rich islands varies for the two films – with PH510 showing significantly larger size aggregates in comparison to P films. The average grain size in P and PH510 films are 40 and 110 nm, respectively (Table 1). This is further verified by considering the surface roughness of the films. The PH510 film has almost twice the surface roughness on a sub-micron scale which originates from the deeper intervening gaps between the large-sized grains. While these films may have similar disorder on the molecular-scale, the increased size of grains may determine an improved transport process in PH510. The observed extent of granularity is an indicator of the extent of phase-segregation process for the excess PSS, and this is more evident from the detailed transport study. Note that both these films have nearly the same ratio for PEDOT-PSS.

The temperature dependence of normalized conductivity of freestanding films (thickness ~ 15-40 μm) of P and PH510, processed with DMSO, is plotted in figure 2(a). In both cases the room temperature conductivity values, typically between 50-100 S/cm, are sensitive to the processing conditions and thickness. Nevertheless, the temperature dependence of conductivity consistently shows the significant difference between P (stronger) and PH510 (weaker). In case of P films, the conductivity decreases by nearly three orders of magnitude, whereas in PH510 it decreases only by a factor of five, as shown in figure 2(a). Recent investigations have suggested a hopping form, of stretched exponential type $R = R_0 \exp(T_0/T)^p$, for the low temperature transport in PEDOT-PSS films. However, these investigations were limited to T > 60 K, which only marks the onset of sharp change in conductance; suggesting that the transport at T < 60 K is more relevant to determine the role of disorder. To investigate this in detail, the reduced activation energy $[W = \Delta \ln \sigma / \Delta \ln T]$ is plotted as a function of temperature, as shown in figure 2(b) [31]. The *W*-plot slope suggests distinctly different transport regime for PH510. For the P-film, the stretched exponential form is justified (rather than the simple exponential form of fluctuation-induced tunneling models) and the value of exponent 'p' is directly obtained from the negative slope of the graph at low temperature (typically T < 40 K), p ~ 0.46. This suggests that quasi one-dimensional (1D) variable range hopping (VRH) is the dominant transport mechanism in P films. The scenario for quasi-1D VRH is mainly due to the important role played by the highly disordered quasi-1D regions in the system, especially at low temperatures; also indicative of the weak interchain transport [32]. Interestingly, the evidence for this type of quasi-1D transport has been observed in recent high-electric field experiments in PEDOT-PSS [29]. The PH510, however, has a temperature independent W(T) at low temperatures indicating that this system is in the critical regime of the metal-insulator (M-I) transition, as has been observed in several conducting polymers [33]. This is further ascertained by considering the resistivity ratio $R_{4.2K}/R_{300K}$, which is lower by two orders in magnitude in PH510 film. At low temperatures (T < 30 K), the resistivity of PH510 can be described by a power law form, $\rho(T) \sim T^{-\beta}$, with $\beta$ ~ 0.31, distinct from the stretched exponential form for VRH. For the critical regime, the resistivity of disordered systems has been shown to follow a power-law dependence rather than the usual activated behavior, and it is expressed in terms of $\rho(T) = (e^2 p_F / \hbar^2)(k_B T / E_F)^{-\beta}$, where $p_F$ is the Fermi momentum, *e* is the electronic charge and $0.3 < \beta < 1$ [34]. The value of the exponent *β* being close to its lower limit (0.3) is indicative that electron-electron interaction effects may be important for the PH510 system. Furthermore, PH510 being in the critical regime of M-I transition suggests that the extent of delocalization is much larger, that results in enhanced interchain transport.

It is well known that in low-dimensional disordered systems, the temperature dependence of conductivity by itself is not adequate to fully understand the mechanism of charge transport; and the complimentary magnetoresistance (MR) data is often required to determine the nanoscopic scale transport parameters. The MR of both systems at low temperatures is compared in figure 3. There is a significant positive MR for P–films [figure 3(b)], but this is considerably less in case of PH510 film [figure 3(a), and Table 1]. The positive MR has a quadratic dependence at low-fields and a tendency towards saturation at high-fields, suggests an origin in the wavefunction shrinkage mechanism under magnetic field. Further, this effect is more pronounced at lower temperatures. The localization length $\xi$ can be estimated from the low-field data for P-film, using the following expression valid for 1D Mott type VRH, $\ln[\rho(B)/\rho(0)] = K(T_0/T)^y (eB\xi^2/\hbar)^2$, where $K = 0.0015$ is a numerical coefficient and $y = 3/2$ [35]. A quadratic fit to the low-field MR data at 1.3 K gives a value for the localization length, $\xi \sim 9\text{-}10$ nm, wherein the value of $T_0$ ($\sim 275$ K) from the temperature dependence of resistance has been used to estimate the above value. This value of localization length favors a transport across conducting regions limited by insulating PSS lamellae in a matrix comprising both the regions, as also indicated by recent works [29, 36, 37]. Due to the carriers being sufficiently delocalized even at low temperatures, any shrinkage of the overlap of the electronic wavefunctions under magnetic-field for the PH510 film does not have a discernable impact on the resistance; whereas in case of P even minor shrinkage of the overlap of wavefunctions can dramatically decrease the conductivity of the system. The absence of a microscopic theory for magnetotransport in the critical regime, however, prevents a precise calculation of the length scale of the delocalized state. This sharp contrast among the two systems has interesting implications for understanding the transport, especially in large-grained film (PH510), and this is described in detail in the section below.

In figure 4, the normalized conductance of two PEDOT-PSS films is plotted as a function of frequency at various temperatures. The frequency dependence of conductance as a function of temperature simultaneously considers the intra-grain and inter-grain processes in the system. At low frequencies, the conductance is nearly constant and it increases as $\sim \omega^s$ ($s < 1$) above a certain threshold frequency, denoted as the *onset frequency* $\omega_0$. This frequency response is characteristic of many other disordered systems. Whereas the $\omega^s$ form holds approximately for the PEDOT-PSS system, the *onset frequency* $\omega_0$ is quite a useful parameter to probe the extent of nanoscale 'connectivity' in the films. A correlation length $\lambda$ can be defined for any disordered system and this length scale corresponds to the distance between connections (e.g. junctions, nodes, etc.) in the system. The charge carrier travels a distance L at a frequency $\omega_L$ and this distance traveled is lower at higher frequencies. At the onset frequency $\omega_0$, the carrier travels the distance $\sim \lambda$. Below this frequency the carriers have to travel across the conducting clusters, which are separated by the insulating PSS lamellae. The onset frequency therefore separates the two regimes of transport [38]. A higher value of onset frequency implies smaller correlation length and shorter connections in the system [38]. For our system, we define the onset frequency as the value where $\sigma(\omega_0) = 1.1\sigma(0)$, where $\sigma(0)$ is the dc conductivity. The PH510 films have superior transport characteristics in comparison to P-films at room temperature as seen from the significantly higher value of the onset frequency (see Table 1). The onset frequency of PH510 films changes only by a small factor, in sharp contrast to the P-films [see figure 4(c)]. Moreover, the increase in conductance as a function of frequency, at all temperatures, is rather weak (well below a factor of 2) in case of PH510; whereas it increases considerably for P samples. The onset frequency for PH510 is nearly temperature independent, whereas in P type it is strongly temperature dependent, as in figure 4(c). The presence of 'shorter network connections' together with a very weak temperature dependence down to $\sim 5$ K, suggest that the limitation on transport in PH510 arises from the connectivity within the PEDOT-rich grain rather than transport via the PSS barriers. This implies that the PH510 system can be electrically modeled as comprising closely packed junctions and nodes, determined by the intrinsic disorder of the *conducting regions*. Such a model also explains the rather weak temperature variation of frequency dependent transport. On contrary, if the transport was insulating barrier-determined a significant frequency dependent transport could have been observed. Whereas in P-films, the strong frequency dependence of ac conductance indicates a significant decrease in the onset frequency $\omega_0$, as the temperature is lowered [see figure 4(b)], indicating the weakening of the temperature-activated transport across the insulating PSS lamellae intervening the conducting PEDOT-rich grains [29]. This difference in behavior, considered together with the magnetotransport data, points to the following surprising result for transport in PEDOT-PSS systems. In both PEDOT-PSS systems, the carriers are confined to *at least* within the span of a single PEDOT-rich domain. In the case of small grain size systems (P), the conduction is limited by the insulating PSS barriers, with the 'connections' weakening with decrease in temperature. At lower temperatures, the phonon-mediated process between the conducting regions gets hampered. However, in the case of PH510, no such weakening of network connectivity is observed since the intervening barriers are sufficiently thin enough and no longer inhibits carrier motion, even at very low temperatures. Therefore, it is the intrinsic disorder in the conducting domains that ultimately limits the conduction in PH510 system. A comparison of the morphology and transport parameters for these two types of conducting PEDOT systems is shown in Table 1.

## 4. Conclusion

In summary, the experimental investigation of the charge transport in thin films of poly(3,4-ethylenedioxythiophene)–polystyrenesulfonic acid (PEDOT-PSS) reveals significant correlation with the morphology of the system. This work compares the magnetotransport and temperature dependent ac-transport parameters across different conducting grades of PEDOT-PSS processed with dimethyl sulfoxide (DMSO). Depending on the subtle details of morphology, the transport at low temperatures is shown to vary from the hopping regime (P) to critical regime of the metal-insulator transition (PH510). It is shown that the conduction mechanism in PEDOT-PSS thin films depends strongly on the extent of aggregation of the constituent polymer nanoparticles. From the magnetotransport analysis, it is shown that the characteristic length-scale for transport correlate with the PSS-barrier limited picture. The onset frequency for the increase in ac conduction as well as its temperature dependence is a tool that directly probes the extent of connectivity in the network; and this data is a confirmation that distinctly different conduction mechanisms are present, depending on the morphology. The evidence from ac transport indicates that the limitation on conductivity arises from the insulating PSS barriers in one system (P), and from the intrinsic disorder present within the conducting regions in the other system (PH510). Whereas both the conduction processes are consistent within the phase-segregation model, the extent of such segregation determines different conduction processes. The results suggest that the present development of the enhancement in room-temperature conductivity in PEDOT-PSS, relying on segregating and lowering the excess insulating PSS barriers, is already close to an optimum scenario, and the future enhancements in PEDOT conductivity must instead address to reduce the intrinsic disorder within PEDOT-rich highly conducting domains.

**Acknowledgements**

The magnetic-field measurements were performed at the DST National facility for Low Temperature and High Magnetic Field, Bangalore (we thank Dr. V.Prasad). S.S.C.S. and M.J. thank CSIR, New Delhi for financial assistance.

**Figures:**

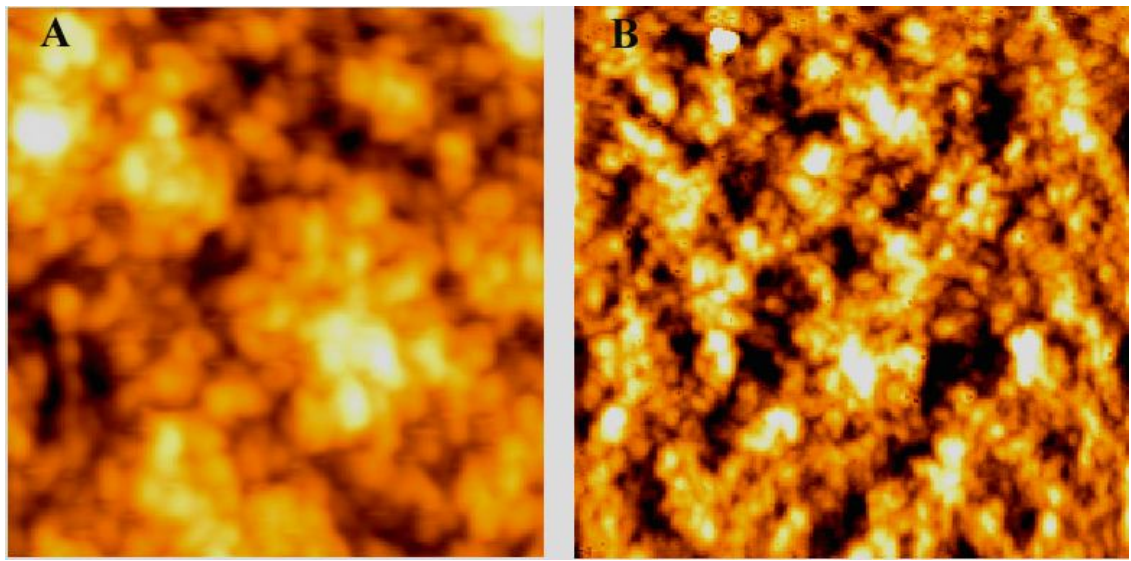

Figure 1. Topography images of PEDOT- PSS film surface obtained with tapping-mode AFM on a scale 2×2 μm$^2$ - (A) PH510 film surface (vertical bar is ~ 20 nm ) (B) P film surface (vertical bar is ~ 10 nm).

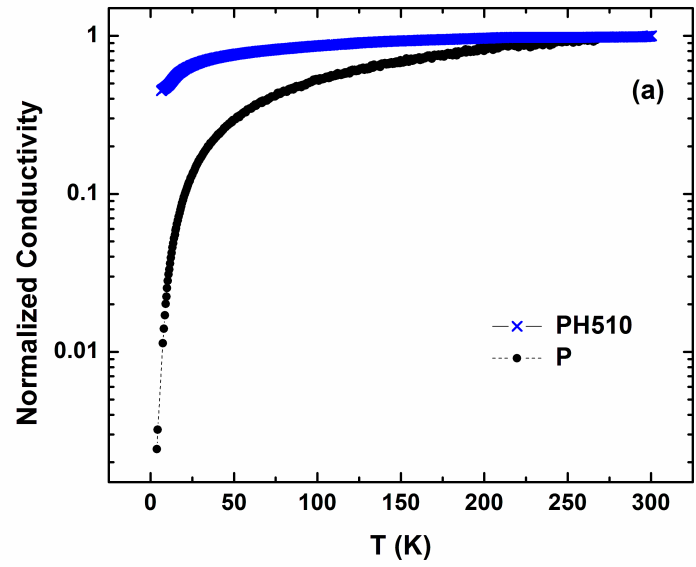

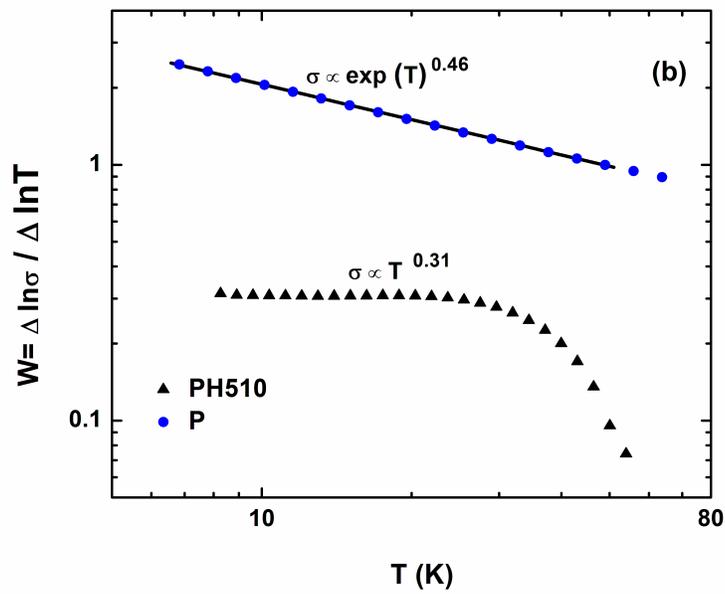

Figure 2. (a) Temperature dependence of normalized conductivity for PH510 and P- films. (b) Reduced activation energy, W (T) vs. T in log-log scale.

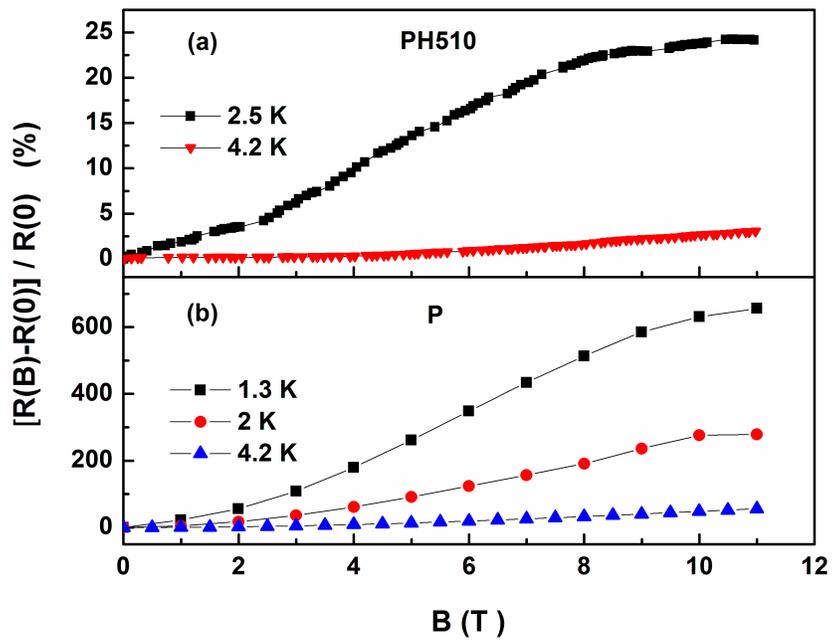

Figure 3. Magnetoresistance vs. field for (a) PH510 (b) P-films.

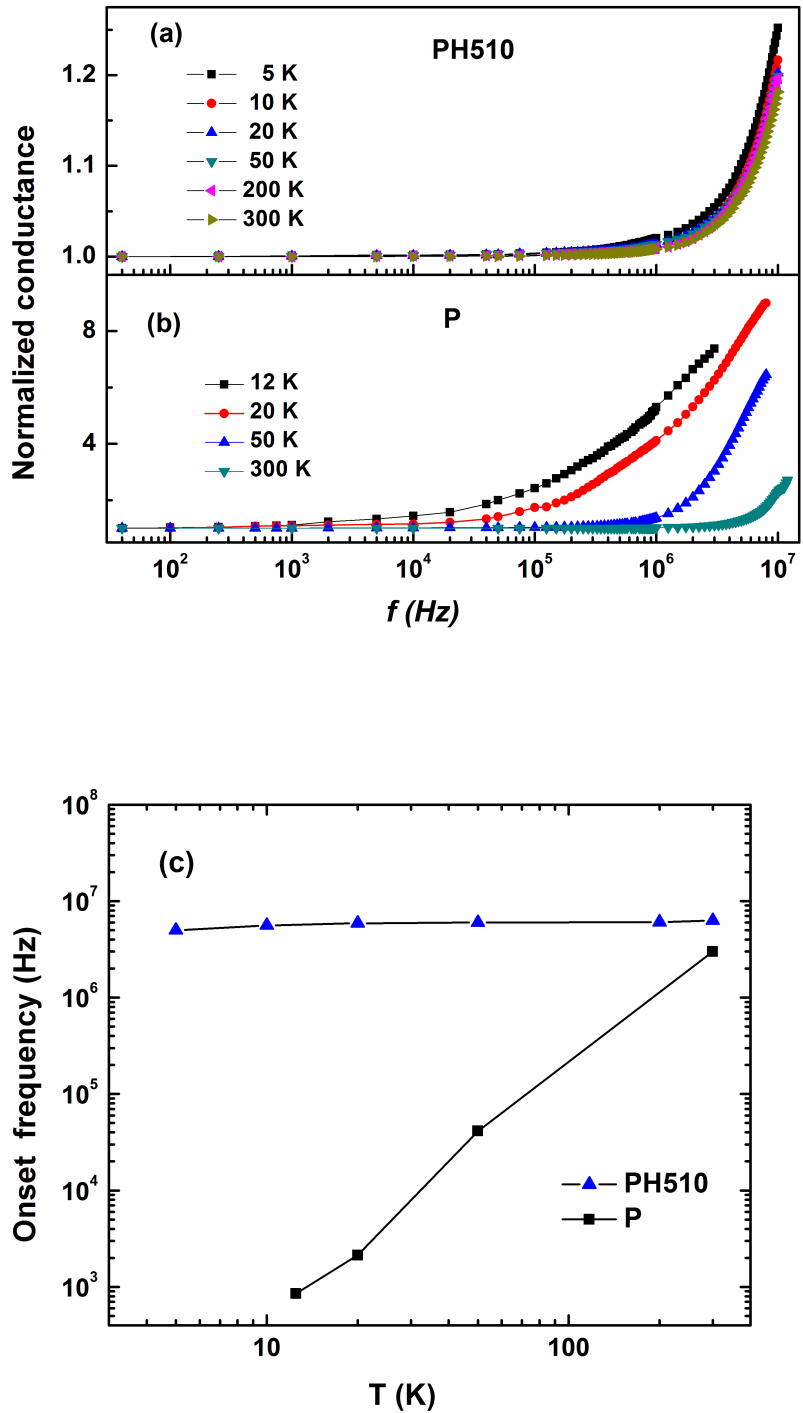

Figure 4. Frequency dependence of conductivity at various temperatures for (a) PH510 -film (b) P- film; (c) Onset frequency vs. T for PH510 and P- films.

Table 1. Comparison of transport parameters for P and PH510 PEDOT-PSS films

| PEDOT-PSS | Average grain size (nm) | Low temperature transport (T < 40 K) | $R_{4.2K}/R_{300K}$ | $\Delta R(B)_{sat.}/R$ (%) at 4.2 K [a] | Onset frequency (MHz) | |
|---|---|---|---|---|---|---|
| | | | | | 300 K | 10 K |
| Baytron P | 40 | 1-D VRH | 285 | 56 | 3 | $8.5 \times 10^{-4}$ |
| Baytron PH510 | 110 | critical regime | 2.8 | 3 | 6.3 | 5.6 |

(a) $\Delta R(B)_{sat}$ denotes the saturation value of MR, at 11 T